\documentclass{pasa}%

\usepackage{threeparttable}
\usepackage{longtable}
\usepackage{multicol}
\usepackage{etaremune}
\usepackage{array}
\newcolumntype{L}[1]{>{\raggedright\let\newline\\\arraybackslash\hspace{0pt}}m{#1}}
\newcolumntype{C}[1]{>{\centering\let\newline\\\arraybackslash\hspace{0pt}}m{#1}}
\newcolumntype{R}[1]{>{\raggedleft\let\newline\\\arraybackslash\hspace{0pt}}m{#1}}

\title[]{{\it Mary}, a pipeline to aid discovery of optical transients}

\author[Andreoni et al.]{I. Andreoni$^{1,2,3,4}$\,$\thanks{igor.andreoni@gmail.com}$, C. Jacobs$^1$, S. Hegarty$^1$,  T. Pritchard$^{1,3}$, J. Cooke$^{1,2,3}$ \and S. Ryder$^4$\\
\affil{$^1$Swinburne University of Technology, PO Box 218, Hawthorn, VIC, Australia, 3122}%
\affil{$^2$ARC Centre of Excellence for All Sky Astrophysics (CAASTRO)}%
\affil{$^3$ARC Centre of Excellence for Gravitational Wave Discovery (OzGrav)}
\affil{$^4$Australian Astronomical Observatory, 105 Delhi Rd, North Ryde, NSW 2113, Australia}}

\jid{PASA}
\doi{10.1017/pas.\the\year.xxx}
\jyear{\the\year}

\usepackage[authoryear]{natbib}
\bibpunct{(}{)}{;}{a}{}{,}
\usepackage{aas_macros}
\usepackage{hyperref} 
\hypersetup{colorlinks,citecolor=blue,linkcolor=blue,urlcolor=blue}

\begin{document}%
\begin{abstract}
The ability to quickly detect transient sources in optical images and trigger multi-wavelength follow up is key for the discovery of fast transients. These include events rare and difficult to detect such as kilonovae, supernova shock breakout, and ``orphan" Gamma-ray Burst afterglows. We present the {\it Mary} pipeline, a (mostly) automated tool to discover transients during high-cadenced observations with the Dark Energy Camera (DECam) at CTIO. The observations are part of the ``Deeper Wider Faster'' program, a multi-facility, multi-wavelength program designed to discover fast transients, including counterparts to Fast Radio Bursts and gravitational waves. Our tests of the {\it Mary} pipeline on DECam images return a false positive rate of $\sim 2.2\%$ and a missed fraction of $\sim 3.4\%$ obtained in less than 2\,minutes, which proves the pipeline to be suitable for rapid and high-quality transient searches. The pipeline can be adapted to search for transients in data obtained with imagers other than DECam.

\end{abstract}
\begin{keywords}
Gravitational waves --
Methods: data analysis -- 
Techniques: image processing -- 
Supernovae: general --
Novae, cataclysmic variables 
\end{keywords}
\maketitle%

\section{Introduction }
\label{sec:intro}
The direct detection of gravitational waves and the design of new facilities to identify fast optical transients open fascinating perspectives for the future of time-domain astronomy. Both large observing programs and small ``boutique" experiments are taking part in the search for electromagnetic counterparts to gravitational wave events discovered by the LIGO/Virgo collaboration \citep[e.g.,][]{Abbott2016b}. These programs can involve large aperture, narrow-field facilities such as the Keck telescopes, as well as smaller telescopes with wide field of view such as the Australian National University's SkyMapper telescope \citep{Keller2007}.   

Fast and faint transients in the nearby Universe are among the most intriguing but elusive events to detect in time-domain surveys. For example, they are considered to be the most promising counterparts to gravitational waves at optical wavelengths. These sources include kilonovae \citep{Tanvir2013,Metzger2016}, gamma-ray bursts - detected as prompt emission, afterglow, or orphans \citep{Vestrand2014, Cenko2015, Ghirlanda2015}, and supernova shock breakouts \citep{Garnavich2016} which are transients difficult to detect because of their expected low-luminosity and/or short time scale. A confident identification and rapid follow up of such transients becomes the key to successful multi-messenger studies \citep{Chu2016}. The most outstanding discoveries in the near future are likely to depend on the specifics and performance of the telescopes involved, as well as the speed of the data analysis. In fact, the prompt identification of interesting fast transients allows rapid follow up that can provide much more valuable information than the detection of the transient alone, even if the transient is detected with a large telescope.

Currently, the Dark Energy Camera \citep[DECam;][]{Flaugher2015} is one of the best imagers in the southern hemisphere to catch fast and faint optical transients. DECam combines a $\sim 3$\,deg$^2$ field of view per pointing with the depth reachable with the 4\,m Blanco telescope at CTIO, allowing to probe extremely large volumes, being sensitive to extragalactic fast transients out to redshift $\sim$\,2. As such, DECam is an excellent instrument to perform observations aimed at detecting optical counterparts to Fast Radio Bursts \citep{Petroff2017}, or counterparts to gravitational wave events in response to LIGO triggers \citep{Cowperthwaite2016}. Its potential to detect kilonovae \citep{Doctor2016}, supernova shock breakout \citep{Forster2016}, and even new classes of fast transients makes DECam suitable also to initiate ``reverse" searches in LIGO data for gravitational wave signals at lower significance \citep{Was2012}.   

Several pipelines have been designed to detect optical transients, usually focusing on objects evolving on timescales of a few days to months. Such pipelines are usually designed for surveys such as, among others, the Supernova Legacy Survey \citep{Astier2006}, the Dark Energy Survey \citep{Kessler2015}, SkyMapper \citep{Keller2007}, the Palomar Transient Factory \citep{Law2009}, Pan-STARRS \citep{Rest2005}, and the Catalina Sky Survey \citep{Djorgovski2010}. These programs unveiled many new classes of Galactic and extragalactic events by regularly surveying large swaths of the sky. In some cases, pipelines were optimised to allow near real-time analysis and detection of transients within minutes from the data acquisition \citep[e.g.,][]{Perrett2010,Rest2014,Cao2016, Forster2016}, leading to the detection of fast transients such as orphan gamma-ray bursts \citep{Cenko2015}. The development of new pipelines able to process huge amounts of data quickly becomes even more important for programs using the upcoming Zwicky Transient Facility \citep[ZTF,][]{Smith2014,Bellm2014} and the future Large Synoptic Survey Telescope \citep[LSST,][]{Ivezic2008}.
 
In this paper we present {\it Mary}\footnote{The name {\it Mary} is a reference to Mary Shelley, author of the ``Frankenstein" novel; the name was chosen to express the idea of several pieces of code brought together to form the final ``Creature".}, an example of a custom-made pipeline that quickly and effectively analyses optical images to detect transient events. We coded {\it Mary} to search for transients in near real time with DECam, in the framework of the ``Deeper Wider Faster" program. We give a brief overview of the ``Deeper Wider Faster" program in Sec.\,\ref{sec:DWF}, we describe the {\it Mary} code in Sec.\,\ref{sec:code} and we present the tests performed on DECam images in Sec.\,\ref{sec:DECam}. A discussion of the parameters choices, the tests results, the possible application of {\it Mary} to data from other optical telescopes, and concluding remarks (Sec\,\ref{sec:discussion}-\ref{sec:conclusion}) complete the paper.

\section{The ``Deeper Wider Faster" program}
\label{sec:DWF}

The ``Deeper Wider Faster" program \citep[DWF, Cooke et al., in prep;][]{Andreoni2017b, Andreoni2017a,Vohl-accepted} organises and conducts several components to detect and study fast transients. 

The first component coordinates {\it simultaneous} observations of the same targeted field, among a list of target fields, with multiple facilities operating at different wavelengths. In February 2017, DWF included simultaneous observations with the DECam optical imager, the Parkes, Molonglo, and Australia Telescope Compact Array radio telescopes, the Rapid Eye Mount optical/infrared telescope in Chile, and the optical/UV/X--Ray/Gamma ray telescopes mounted on the NASA {\it Swift}\footnote{swift.gsfc.nasa.gov} satellite. The observing strategy with DECam is based on fast (20\,s exposure time), continuous sampling of successive target fields for 1-3 hours each per night. The fields are chosen depending on their visibility from both Chile and Australia and are observed for multiple consecutive nights, usually 6 per semester. 

The second component consists of the near real-time data analysis and candidate assessment that takes place in the collaborative workspace environment organised for DWF at Melbourne University or at the Swinburne University of Technology \citep{Meade2017}. 

The third component of DWF consists of rapid response and conventional Target of Opportunity (ToO, imaging and spectroscopy) follow up. While some of these facilities can follow up transients only a few hours after their identification due to the geographical location of the observatories, the Gemini South 8\,m telescope can perform rapid spectroscopy minutes after significant discoveries due to its proximity to the Blanco telescope, as well as its rapid response Target-of-Opportunity program. Classically scheduled follow-up observations (imaging and spectroscopy) constitute the fourth component of the DWF program.

Fast and effective data analysis is crucial to catch fast transients and successfully obtain rapid response, ``flash'' spectroscopy, which can add valuable information to understand the mechanisms at the basis of supernova explosions \citep{Gal-Yam2014}, including optical shock breakout. It also offers our only opportunity to study and characterise new exotic fast transients. DWF organised a network of small to large telescopes to follow up the transients discovered with the {\it Mary} code, including the Anglo-Australian Telescope (AAT), Gemini South, the Southern African Large Telescope (SALT), and the SkyMapper and Zadko telescopes \citep{Coward2016} in Australia.   

During DWF campaigns, {\it Mary} automatically generates lists of highly significant candidates, providing information that populates the database described in Sec.\,\ref{secsub: products for the visualisation}. Experts validate and further prioritise the candidates using light curves, catalog information, and products for the visualisation generated with {\it Mary}. This process leads to manual triggers of large, narrow-field facilities to follow up the most interesting events.

\begin{figure*}
%\begin{center}
\includegraphics[width=2\columnwidth]{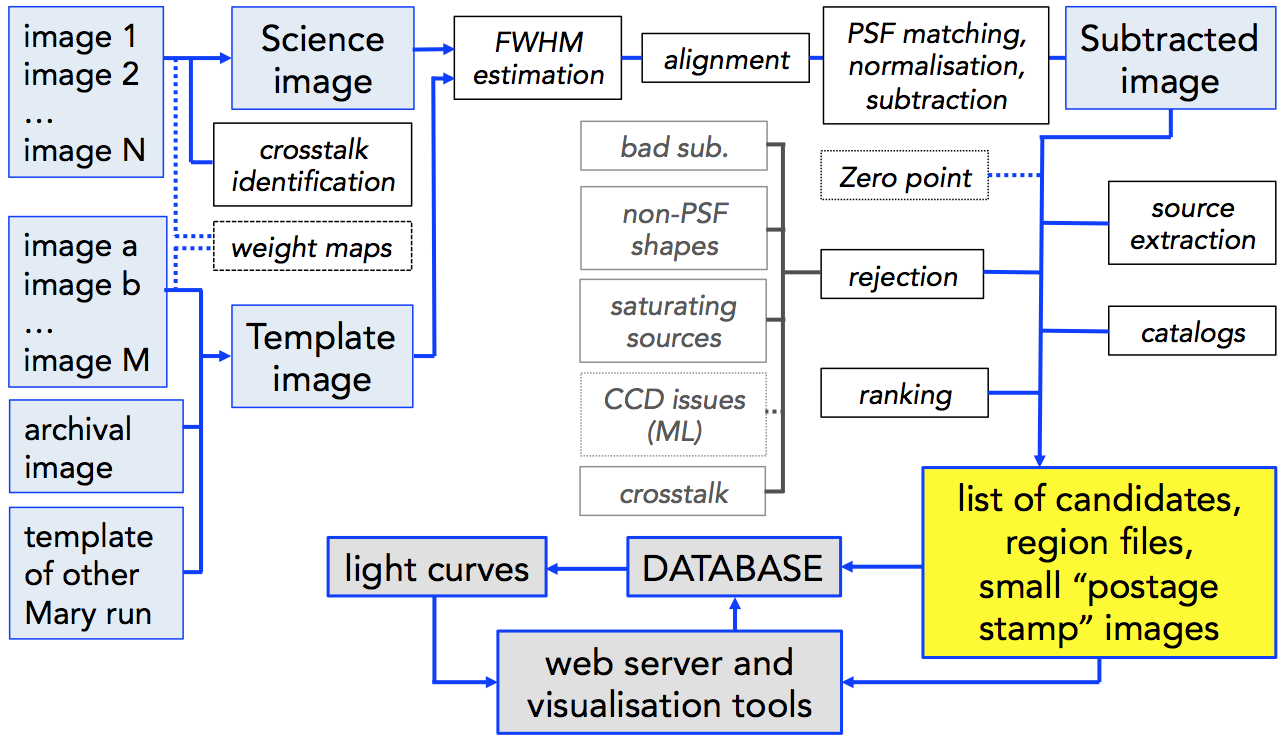}%
\caption{The {\it Mary} workflow, highlighting the main steps leading to the identification of transient events. Dashed lines indicate optional steps that the user can activate. All the indicated operations happen in parallel for each of the 60 functioning CCDs of DECam. The main steps for the selection of good candidates include the rejection of those sources which are badly subtracted, non-PSF shaped, saturating, possibly associated with crosstalk effects, or flagged as possible CCD artifacts by the machine learning (ML) classifier. The storage of the information in a database, the generation of the light curves, and the display of the products for the visualisation take place outside the parallel process, thus we marked them with a grey background. } 
%\end{center}
\label{workflow}
\end{figure*}

\section{The {\it Mary} pipeline}
\label{sec:code}

We wrote the {\it Mary} pipeline to serve as a fast, simple and flexible tool to analyse the 60 CCD mosaic of DECam in near real time (seconds to minutes after the data are acquired) during DWF observations. The pipeline is also suitable to perform accurate late-time analysis on hundreds of images, with the goal of discovering transients of any type, evolving at nearly any timescale and with nearly any trend in their light curve. However, the main focus of the DWF program is to unveil fast transients, especially the least understood ones. The pipeline is adaptable to telescopes/imagers other than Blanco/DECam, as we discuss in Sec.\,\ref{subsec:applications}. 

The lack of a well defined ``training set'' at the beginning of the novel DWF project forced us to think of a new way to distinguish real transients from cosmic rays and bogus detections without adopting machine learning techniques (Sec.\,\ref{secsub:selection of the candidates}). A machine learning step was introduced in February 2017 to reject a specific type of CCD artifact, as described in Sec.\,\ref{subsec: ML}. We present how {\it Mary} prioritises the candidates in Sec.\,\ref{secsub:catalogs and ranking} and the results of timing and efficiency tests in Sec.\,\ref{subsec: timing test}-\ref{tab: completeness efficiency}. Finally we compare the results with the performance achieved by some of the best pipelines in Sec.\,\ref{subsec:comparison}. In the Appendix we report a table of the parameters - set automatically or by the user - that fully regulate the analysis that {\it Mary} performs.

{\it Mary} is coded in the IDL programming language and runs on CPU machines on the Green II supercomputer at Swinburne University of Technology. It ties together popular packages such as the AstrOmatic\footnote{http://www.astromatic.net/} packages Swarp and SExtractor \citep{Bertin1996,Bertin2002}, HOTPANTS \citep{Becker2015}, and new custom programs that facilitate the work flow and the specific selection of candidate transient sources. In this paper, we assume that the images have already been astrometrically calibrated and pre-processed with bias and flat field corrections. We will discuss individually the main stages driving the workflow (shown in its most basic structure in Fig. \ref{workflow}) of the pipeline:  the preparation of a ``template" image to use as a reference and a ``science" image, the image subtraction, the selection of the candidates, and the creation of products to facilitate the visual inspection.

%\twocolumn

\subsection{``Template" and ``science" images}
\label{secsub:science and ref images}
{\it Mary} takes advantage of image subtraction techniques to identify sources changing their flux between different epochs. The science image is obtained coadding a number ``N" of calibrated images (Fig.\,\ref{workflow}), or using an individual image when only one is available or during the search for fast transients. The template image can be obtained in three ways:

\begin{itemize}

\item using a deep archival template image, well separated in time with respect to the epoch of the observation, depending on the timescale at which the target transients evolve. This method usually allows the discovery of the largest number of transients.

\item coadding a number ``M" of individual calibrated images. This becomes particularly useful when an ``old" template image is not available, for example during the follow up of a gravitational wave trigger in a region of the sky not previously covered.

\item copying the template from a previous {\it Mary} run. This method becomes handy when the analysis is performed in near real time with continuous imaging of the same field, as it sensibly reduces the {\it Mary} running time. The time saved depends on the number of images to coadded when the user chooses to avoid copying the template.

\end{itemize} 

{\it Mary} coadds the input images using the Swarp package. The user can switch on/off the optional automatic creation and usage of weight maps. After the creation of the science and template images, {\it Mary} automatically estimates the seeing FWHM of the images in the following way:
\begin{itemize}
\item running SExtractor on the images;
\item selecting those sources with a Star/Galaxy classification ranking $>0.98$. If less than three sources fulfil this requirement, sources with a ranking $>0.96$ are selected;
\item calculating the average FWHM\_IMAGE that SExtractor outputs for the selected sources.
  
\end{itemize} 

The automatic estimation of the seeing FWHM allows the definition of other parameters that, otherwise, would have to be set manually by the user for each set of images to search with {\it Mary}: those parameters are among those marked with the letter ``A" in Table\,\ref{tabpar}. Also, this module prepares a list of bright and saturating sources that can be removed from the final list of candidates at later stages (see Sec.\,\ref{secsub:selection of the candidates}).

When science and template images are ready to use, they are aligned in image coordinates using the resampling functions of Swarp. Usually, it is convenient to ``frame" the common area between science and the template images (using the {\it xdim, ydim, fixoldchoice} parameters; see Appendix A) to allow i) a proper alignment in image coordinates; and ii) a more polished image subtraction, thanks to a maximisation of the common sky area over the non-overlapping area. This framing operation is not needed when the template image is the coaddition of many largely-dithered images, covering the whole area of the science image. During DWF observations we rarely have such images available, because even the best templates available leave ``stripes'' (a few pixels wide) of the new science images without coverage. Therefore, knowing the shifts occurring during the observations and the precision of the astrometric calibration, the pipeline can be set up to different sizes for the final aligned images. The optimal size usually adopted in near real time and during the tests presented in this paper is 3800$\times$1800\,pixel, while for other works we adopt 4000$\times$2000\,pixel or the exact dimensions of the CCD. The aligned images are kept available for the user to be displayed with region files overplotted (Sec.\,\ref{secsub: products for the visualisation}). 

A bad alignment ($\gtrsim 1.0 \times$\,seeing FWHM) would result in the failure of the image subtraction, or in a non-perfect subtraction that results in an error message, usually triggered by the unexpectedly large number of candidates exceeding the {\it maxsources} number set in the parameters file (with a default value of 60 per CCD during the near real-time analysis). The images are stored rather than deleted as temporary files, in order to be checked at any time for quality assessment and to facilitate the generation of light curves with programs that we coded separately to the {\it Mary} pipeline.

\subsection{Image subtraction}
\label{secsub:subtraction}

When science and template images are aligned in physical coordinates, {\it Mary} uses the HOTPANTS package to perform the main steps of image subtraction, such as sky subtraction, normalisation of the flux and PSF matching. {\it Mary} computes input parameters such as the average sky level and its standard deviation. The user can indicate the number of standard deviations of the sky to set the threshold below which pixels are considered ``bad pixels" to be masked. Similarly, the user can set the maximum pixel value; otherwise HOTPANTS extracts this information from the header of the image (``SATURATION" keyword). Finally, the user can switch on/off the creation of data quality masks with HOTPANTS depending on the minimum and maximum pixel values to limit the effect of saturating sources, bad columns and bad pixels. An optional step (activated with the {\it photchoice} parameter) determines the zero point of the template image by measuring the magnitude of a number (usually 30-50 per CCD) of pre-selected stars present in the target field and comparing them with the values reported in existing catalogs. Specifically, we use the USNO-B1 or Gaia catalogs (determined with the {\it catalogcalib} parameter), as the Sloan Digital Sky Survey is largely incomplete in the southern hemisphere.     

It may be possible to implement the optimal subtraction algorithm \citep{Zackay2016} in future versions of the {\it Mary} code, with the intent to compare the results against those presented in this paper.   

\subsection{Selection of the candidates}
\label{secsub:selection of the candidates}

Sources varying their luminosity between the template and the science images leave residuals on the images generated by the subtraction with HOTPANTS. We run the SExtractor package on the image resulting from the subtraction to detect and measure such residuals. The user controls the main extraction and photometry parameters (DETECT$\_$MINAREA, THRESH$\_$TYPE, DETECT$\_$THRESH, ANALYSIS$\_$THRESH) from the {\it Mary} parameter file. The result consists of a catalog of a (typically) large number of sources, dominated by non-astrophysical detections. 

\begin{figure*}
\begin{center}

\includegraphics[width=2\columnwidth]{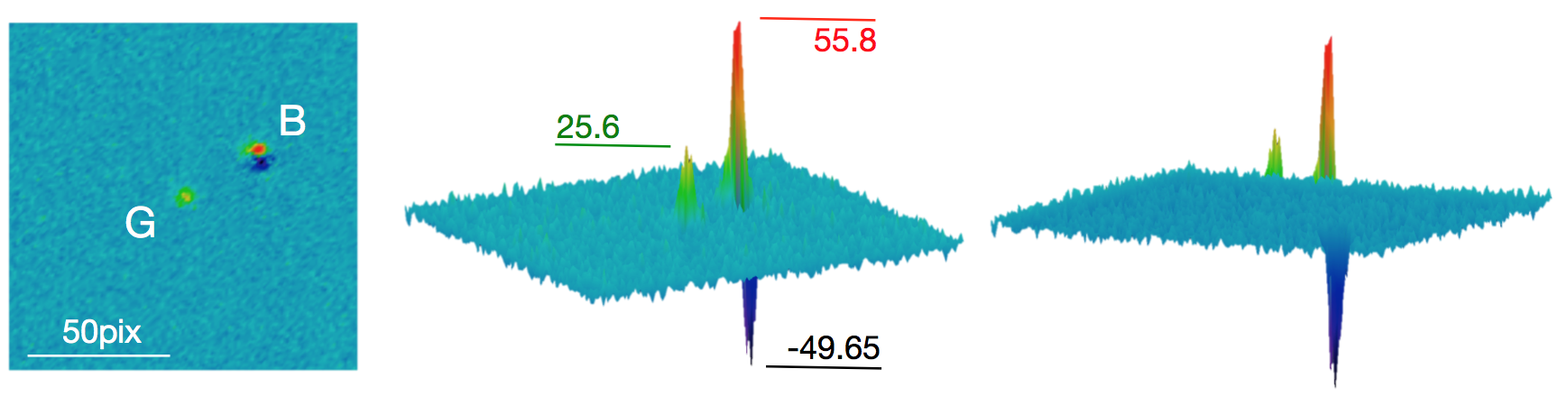}%
\caption{Example of image subtraction leaving ``good" or ``bad" residuals, indicated with the letters G and B in the left figure, respectively. The figures at the centre and at the right present a 3D rendering \citep[generated with the AAOGlimpse software,][]{Shortridge2012} of the result of the image subtraction, from two different viewing angles. The G residual lacks the negative peak associated with the B residual. ADU values of the positive and negative peaks are indicated in the central figure.”}
\label{residual}
\end{center}
\end{figure*}

We first apply a number of criteria to pre-select candidates, for the specific instrumental parameters of DECam and the seeing FWHM at CTIO, such as a shape elongation upper limit, a minimum isophotal area, and minimum FWHM value for the source (regulated by the {\it elmax, isoareafmin, fwhm} parameters, see Appendix). The default values we set during the DWF observations are: {\it elmax}=1.8, {\it isoareafmin}=20, and a minimum FWHM value for the residual equal to half the {\it fwhm} parameter associated with the subtracted image. This initial step allows the rejection of most of the remaining cosmic rays (especially when the science image is created without median-stacking $\geq3$ images) and highly asymmetric artifacts. In fact, bright and saturated point sources rarely subtract properly and usually leave a ``negative'' flux signature adjacent to a ``positive'' flux excess (Fig. \,\ref{residual}). Data quality masks generated with HOTPANTS (see Sec. \ref{secsub:subtraction}) help reject such bogus sources. In addition, {\it Mary} creates an additional catalog on the inverse (i.e., taking $-1.0\,\times$\,\small{VALUE} per pixel) of the subtraction products and, after matching the two catalogs, rejects all candidates associated with detection in the ``negative'' image. After these selection steps, more than $99\%$ of bogus sources are rejected without the need of machine learning, when the images are aligned within $\sim$ the FWHM of the image with poorer seeing. Completeness and efficiency tests of these selection methods are reported in Sec.\,\ref{subsec: tests}.

In the specific case of the double-amplifier CCDs of DECam, electronic crosstalk\footnote{See the DECam known problems web page: \url{http://www.ctio.noao.edu/noao/node/2630}} associated with bright point sources can cause ``ghost" sources to appear at a location on the CCD symmetrical with respect to the contact line between the two amplifiers. Crosstalk features are often similar in appearance to real point sources. Before the coaddition of the images, {\it Mary} identifies bright sources (saturating or brighter than a user-defined threshold) and flags the regions of the CCD where crosstalk effects are expected.

%\onecolumn

%\twocolumn

\subsection{Machine Learning classifier to reject CCD artifacts}
\label{subsec: ML}

\begin{figure}
\centering
\includegraphics[width=\columnwidth]{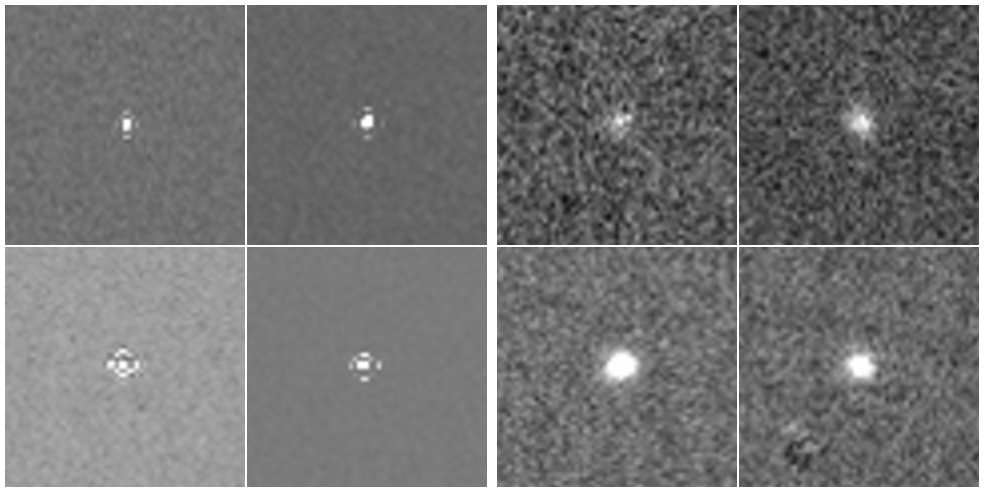}
\caption{Examples of artifacts ({\it left}) and possible good candidates ({\it right}) present in the training set used for machine learning. Our training set
consisted of 559 false positives and 524 potential transients selected
by visual inspection. Only the central 16x16 pixel area was
considered to train the machine learning algorithm.}\label{fig:training_set}
\end{figure}

In order to further purify our candidate sample, we employ machine
learning to reject bad pixels and CCD artifacts, such as those shown in in the left panel of Fig.\,\ref{fig:training_set}.
Machine learning techniques take a data-driven approach to
problems such as the classification and clustering of data in various parameter spaces, instead of
relying on human intuition regarding the significant features in the
data \citep[see][for an overview]{Jordan2015}. Here we consider a \emph{supervised learning} problem, where we used a pre-prepared, labelled data set to train our algorithm to classify the candidates into one of two categories (candidate or artifact).

To prepare the training set we visually inspected $\sim$1000
candidates from the selection described in Sec.\,\ref{secsub:selection of the candidates} and classified them as ``good'' (possible
transients) and ``artifacts'' (unlikely to be genuine sources). This resulted in a set of 524 good candidates and
559 false positives. We augmented this training set by rotating each image
through three \(90\deg\) rotations, for a total training set of size
4332. Given we are searching for point sources (with typical FWHM=1-2$"$), and we want to keep good
candidates even if they occur close to one of these artifacts, we
restricted our training images to a 16x16 pixel stamp (equivalent to 4.2$''$ on a side) centred on the candidate object. We reserved 20\% of
the training images as a test set, in order to evaluate the performance
of trained algorithms on data not used for training purposes. Some
sample images from our training set are depicted in Fig.\,\ref{fig:training_set}.

We used this training set to train two machine learning methods. As the
inputs are 16x16 pixel areas of a single-band FITS file - 256 floating
point numbers - they are of a dimensionality tractable by most
traditional classification techniques. We first employed a support vector
machine \citep[SVM,][]{Cortes1995}, an efficient classifier able to
distinguish between classes with highly non-linear decision boundaries.
We also applied a convolutional neural network \citep[CNN,][]{Fukushima1980,
Lecun1998}, a specialisation of an artificial neural network
\citep[][for a more recent
review]{Rosenblatt1957, Lecun2015} designed for complex image data and computer vision
applications. We trained the SVM using the scikit-learn \citep{Pedregosa2011} python library. We constructed the CNN with three convolutional
layers and two fully connected layers of 128 neurons each. We trained our
network using the Keras package on an NVidia GTX1070 GPU.

The accuracy, defined as the number of correct classifications divided by the total number of classifications, is presented in Table~\ref{tbl:accuracy}; due to the limited size of the test set we the true misclassification rate could be as high as 4.5\% within a 95\% confidence interval. Despite the small size of the input images, we found that the
convolutional neural network achieved slightly better performance on our test
set of 866 images, divided evenly between the two classes. Although understanding the decision criteria inside the network is non-trivial, the network appears to have a developed a robust representation of the gaussianity of the genuine sources and the less continuous artifacts in the data. The CNN-based classifier was incorporated into the {\it Mary} pipeline. The user can choose to take advantage of this module by setting the {\it mlchoice} parameter and can define the threshold of the classification with the {\it classthresh} parameter.

\hypertarget{tbl:accuracy}{}
\begin{table}
\centering

\caption{\label{tbl:accuracy}Training and test set accuracy for the
machine learning methods, trained on a training set of artifacts and
transient candidates. SVM = Support Vector Machine; CNN = Convolutional
Neural Network. We incorporated the CNN-based classifier into the {\it Mary} pipeline.}

\begin{tabular}{@{}lll@{}}
%\toprule
\hline
\hline
\ & Training accuracy & Test set accuracy \\   
\hline
SVM & 98.11\% & 97.21\% \\
CNN & 97.96\% & 98.79\% \\
\hline
\hline
%\bottomrule
\end{tabular}

\end{table}

For example, during the February 2-7 2017 DWF run (see Cooke et al. in prep, for more details), the CNN-based classifier classified 24153/89491
(27\%) of candidates shown as artifacts, leaving 73\% to undergo further automatic selection with {\it Mary} and finally be inspected by
project astronomers and volunteers. At this time we have not performed a
systematic inspection of all rejected candidates, but given the
conservative settings employed and current analysis we expect false positives to be very low. Performance of the machine learning step can be improved with a larger and more diverse training set, allowing better optimisation of the network training parameters. Future development could include an online learning component, where false positives are fed back into the algorithm in real time to allow ``on the fly" retraining.

\subsection{Catalogs query and ranking of the candidates}
\label{secsub:catalogs and ranking}
When large numbers of candidates pass the selection criteria, the user may find it useful to see a priority value attached to each candidate. {\it Mary} assigns the priority values by combining two types of information: those retrieved by querying online or downloaded catalogs, and information about the number of times that a transient was detected. In fact, transient searches usually imply target fields to be observed several times with a defined cadence; thus knowing if, when, and how many times a transient was discovered in the past  greatly helps to understand the nature of the detected sources. During DWF observing runs aimed at discovering extragalactic transients, we set the Guide Star Catalog 2.3 \citep[GSC,][]{Lasker2008} as preferred catalog for {\it Mary} to query, as it provides a basic star/non-star classifier for objects in the southern hemisphere down to relatively faint ($>20$) magnitudes. The search radius (expressed in arcsec) is defined as {\it radiusGSC}=$\sqrt{seeing^2 + u^2}$, where the seeing is defined as {\it fwhm}\,$\times$\,{\it scale} and {\it u} is the angular resolution of the GSC catalogue, $u=1''$. 

{\it Mary} prioritises the detected sources by assigning a score from 1 to 5, where higher value indicates a higher priority for the source to be followed up. The ranking system is optimised for the typical $\sim$6 consecutive observing nights of DWF, and it works as follows:
%\begin{enumerate}
\begin{etaremune}

\item source detected only in the current observing night and not reported in the GSC catalog. 
\item source detected only in the current observing night and reported as possible non-star in the GSC catalog;
\item source detected in the current observing night and in the previous observing night, not reported in the GSC catalog or reported as possible non-star; 
\item source detected in the current observing night and in two or more previous observing nights, not reported in the GSC catalog or reported as possible non-star; 
\item source classified as ``star" in the GSC catalog;
\end{etaremune}
%\end{enumerate}

The query of the GSC catalogue is less effective than the query of catalogs generated on DECam images using the SExtractor star/galaxy classifier, or more sophisticated classifiers such the algorithm described in \cite{Miller2017}. However, the query of online catalogs becomes extremely valuable when searching for transient sources with template images acquired only a few minutes or hours before the science images, for example during the first nights of DWF observations or during the follow up of GW events. The program can be slightly modified to allow the query of different or additional catalogs in the future. 

\subsection{Products for visualisation}
\label{secsub: products for the visualisation}

\begin{figure}
\centering
\includegraphics[width=\columnwidth]{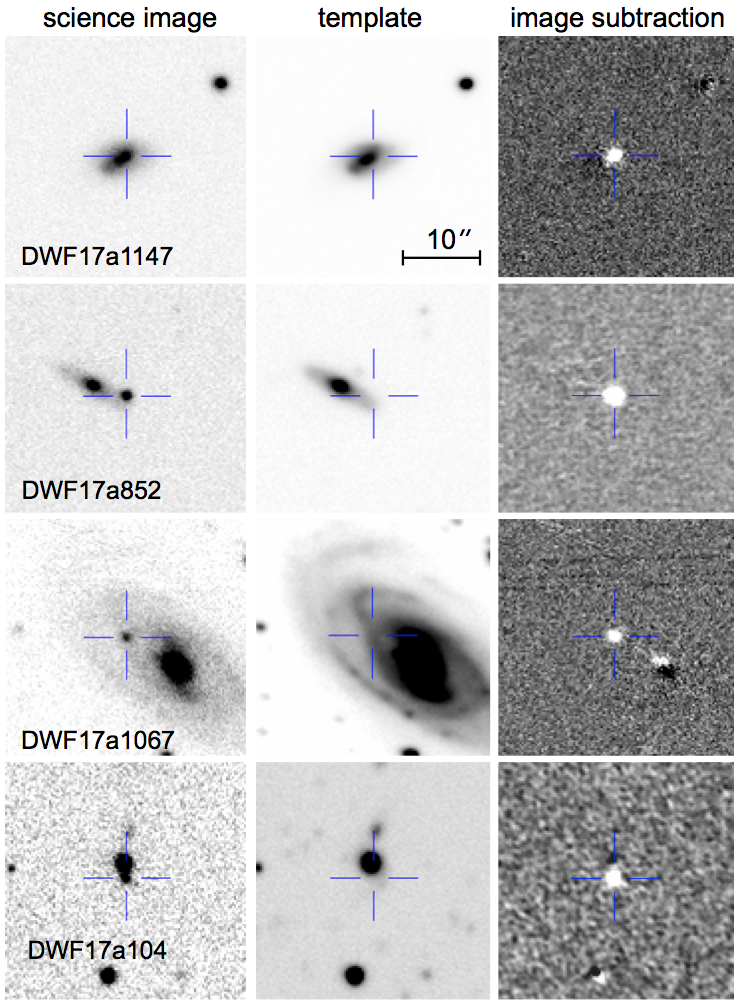}
\caption{Examples of transient sources detected with the {\it Mary} pipeline in near real time. In particular, these images show the small ``postage stamp" images that allowed the identification of the possible superrnovae DWF17a1147, DWF17a852, DWF17a1067, DWF17a104 presented in \cite{Andreoni2017a}. The science images, where the transient is present, and the template images populate the first two columns (in inverse-color grayscale). The image subtraction is presented in the last column for each candidate.}\label{fig:SNe}
\end{figure}

The list of candidate transients comes with a set of products to facilitate the visual inspection of the candidates. Along with the priority assigned to each candidate, the products to help identify and classify interesting sources include:

\begin{itemize}
\item Small ``postage stamp" images with user-defined size cut from the template, the science and the subtracted images, centred on the coordinates of the candidates; some examples are presented in Fig.\,\ref{fig:SNe}. 
\item Region identification files suitable for ``ds9" image visualisation tool with both circles and projection shapes. These files mark the locations of objects on the images, with different colors depending on the object class as identified in the catalogs such as the GSC.
\item Light curves built by collecting magnitude values extracted from the subtracted images (with the same template) created during a series of {\it Mary} runs. These light curves are calibrated if the {\it photochoice} parameter is set, otherwise a zero point = 25 is assumed.
\item Calibrated aperture and PSF photometry light curves. To optimise the fast completion of the transient identification, we consider this part of code detached from the {\it Mary} pipeline (but can be automatically started for each detected source), because the calibration can take minutes of computational time, depending on the number of images available - usually a few hundreds in the DWF program context. 
\end{itemize}

These {\it Mary} products become particularly useful when observations are taken with only a single (or no) filter, as it is not possible to classify the discoveries based on color information, or during searches in near real time. For instance, the DWF program uses images in the $g$ filter for fast transient detection, complemented by color information taken at a much slower cadence.

A custom program consolidates Mary's results into an SQL database on the Green II supercomputer. Once each Mary run is complete, all data pertaining to each identified candidate (including information about its associated image products) is stored to the database. This facilitates simple, powerful data analysis, including straightforward generation of light curves for each observed candidate. The database can be queried from the command line, programatically or via a custom-built web interface, which allows the user to request and view data, candidate ``postage stamps" or compressed CCD images, and to add manual candidate classifications.

We coded additional custom programs to cluster the detections into groups of interest or to better characterise them, for example building well-calibrated light curves using all the images available, by ``unfolding'' the images coadded to generate the science images during one or multiple {\it Mary} runs. These additional programs and tools are not discussed in this paper, because their application strongly depends on the type of observations performed and lies outside the essential structure of Mary.

\section{Tests on DECam images}
\label{sec:DECam}

The {\it Mary} pipeline is designed to work both in near real time during the observations, as well as at later times. We developed {\it Mary} independently from other pipelines built to analyse DECam data, such as PhotPipe \citep{Rest2005, Rest2014}, DiffImg \citep{Kessler2015}, or pipelines designed for other transients projects \citep[e.g,.][]{Forster2016}. {\it Mary} is designed to detect transients in images already calibrated, thus we established a profitable collaboration to calibrate the DECam images in near real time analysis with part of the PhotPipe reduction code. All the subsequent steps (from the creation of science and template images to the end of the pipeline) do not depend on PhotPipe or any other pipeline. For a more refined and complete late-time analysis we take advantage of images calibrated with the NOAO High-Performance Pipeline System \citep{Valdes2007, Swaters2007}. The main differences between the near real-time and late-time transient detection with {\it Mary} lies in the quality of the pre-processing of the images available and the percentage of CCDs successfully accomplishing the calibration without further intervention.

Fast and efficient analysis of the images was the goal at the outset of the {\it Mary} design, which becomes particularly challenging when dealing with the DECam data products. These consist of a collection of 60 functioning 2k x 4k pixel CCDs (32\,MB each), one of which (S7) we ignore due to calibration problems. For DWF, we analyse each CCD in parallel, using one CPU core of the Green II supercomputer per CCD. 
The individual processes undergo no further parallelisation. We coded a script that manages the submission of the jobs to the Portable Batch System queue of the Swinburne supercomputer: dedicated nodes make the time between the submission and the running of each job negligible ($<$1\,s) during the the DWF observations. The user can switch between the two main architectures by setting the {\it pipesetup} parameter to ``RT" to process in near real time the images calibrated with PhotPipe, or to ``NOAO" to analyse multi-extension images processed with the Community Pipeline. The {\it path\_original} parameter allows a choice of path to find the NOAO processed images, while the products are always organised with the same structure to facilitate their retrieval.

The time needed to complete the data processing depends on the number of images to be coadded, the generation and usage of weight maps, the detection threshold, and the set of parameters chosen for the desired configuration of the pipeline. We tested the pipeline to present an example of timing and completeness estimates in the specific configuration described below.

\subsection{Timing test}
\label{subsec: timing test}

In this section, we tested the timing performance of the {\it Mary} pipeline, without accounting for the data transfer time. The data compression and transfer from CTIO (in Chile) to the Swinburne supercomputer (in Australia) affects the near real-time observations in a way extensively discussed in two separate papers \citep[Cooke et al. in prep;][]{Vohl-accepted} specifically dedicated to the DWF program. In particular, \cite{Vohl-accepted} present the custom compression code based on JPEG2000 standard for the compression of DECam data that significantly reduces the time needed to transfer the images from CTIO to the Swinburne supercomputer in Australia. Improved data transfer speed allows a successful near real-time analysis during fast-cadence time-domain surveys such as DWF, without the need of a supercomputing facility at the telescope.  

\begin{figure}
\begin{center}
\includegraphics[width=\columnwidth]{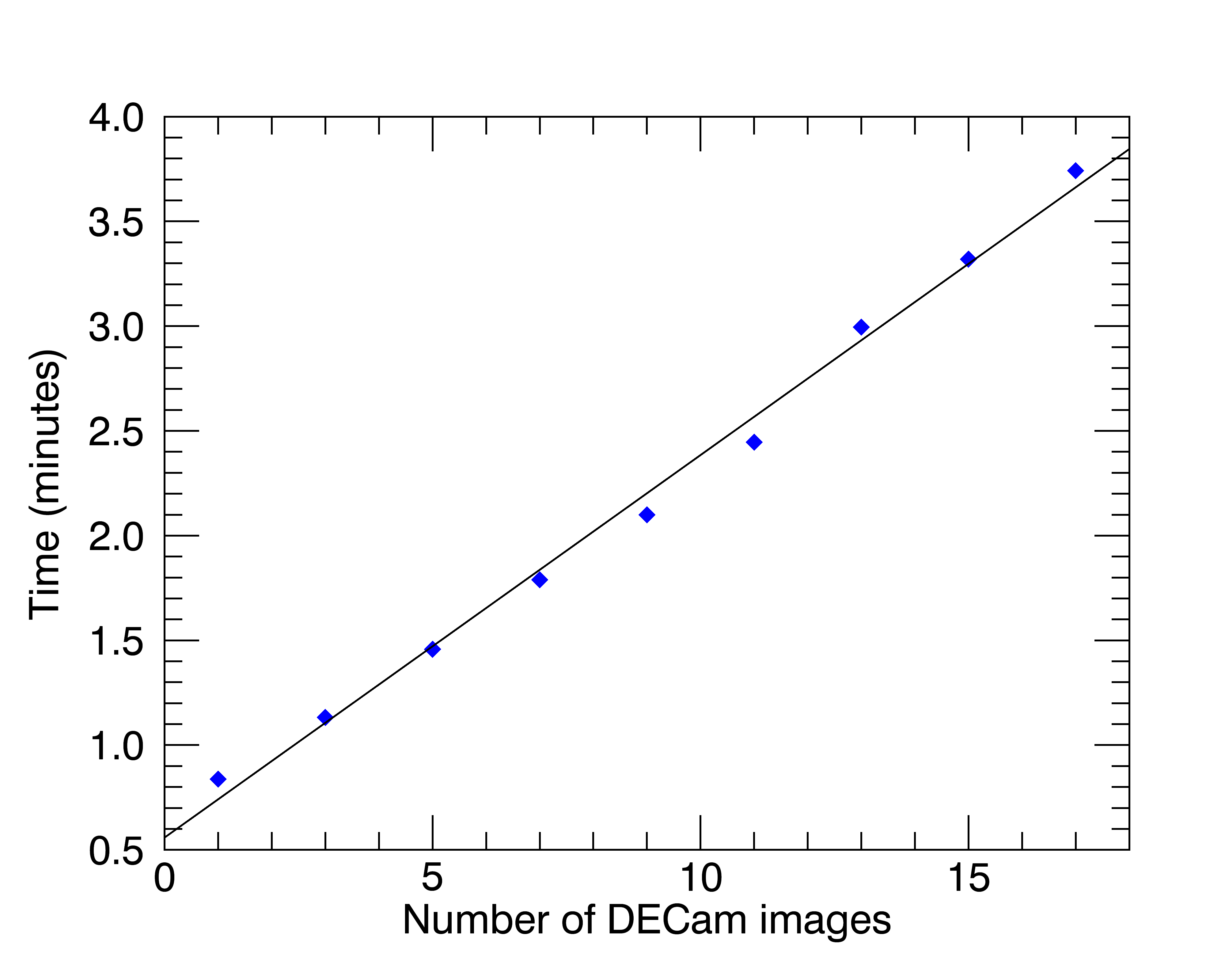}%
\caption{Median time needed to complete all the steps of the pipeline in parallel for 59 functioning CCDs of DECam. We timed the pipeline using different numbers of images to be coadded to form the science image, always using the same 11-month-old template image.}
\label{plot: timing}
\end{center}
\end{figure}

We ran our timing test on images taken on 21 December 2015 UT, during a DWF observing campaign, and calibrated with the NOAO pipeline. The conditions were stable, with seeing around $\sim$1.5$''$ , and we acquired 84 continuous exposures of 20\,s in $g$ band. 
We chose to use self-generated weight maps for the coaddition, and downloaded catalogs to calibrate the zero point and to match the sources with the GSC catalog. We set the following parameters for the source extraction: DETECT$\_$MINAREA=8, DETECT$\_$THRESHOLD=1.8 (for ``positive" images, which corresponds to an effective S/N $\sim 7 \sigma$ for point sources), DETECT$\_$THRESHOLD=1.3 (for ``negative" images). 
The results of the timing test are shown in Fig. \ref{plot: timing}, which shows the average time taken to complete all the operations, from the coaddition of science images (whose number constitutes the horizontal axis) to the building of the products for the visualisation. In order to get a larger sample, each point represents the median of the time values obtained during three runs of the pipeline, keeping exactly the same configuration. A line with slope 
m=0.18 and intercept c=0.56
fits the data quite well, as the $\chi$-squared of the fit returns a p-value=1.00 for 8 degrees of freedom. This means that the probability of the scatter of the points from the line lies within the statistically expected scatter. It is important to note that the crosstalk rejection needed for our dataset strongly affected the timing results, adding about 2-3\,s per individual image. This could be avoided by either a) performing dithered pointings during the observations; b) further improving the crosstalk correction during the calibration; c) running the pipeline on images taken with other cameras not affected by CCD operating system crosstalk effects.

\subsection{Completeness and efficiency tests}
\label{subsec: tests}

We tested the performance of the {\it Mary} pipeline in different conditions of seeing and depth, using 59 CPU cores and 1 GPU of the Swinburne Green II supercomputer. The data we used were taken between 19-22 December 2015 during high-cadenced DWF observations. Seeing conditions, limiting magnitudes, and the results of the tests are summarised in Table\,\ref{tab: completeness efficiency}. We estimated the significance above the background by performing PSF photometry of the sources in the science image, then evaluating an approximate signal-to-noise ratio (S/N) using the formula\footnote{see, for example, \url{http://www.eso.org/~ohainaut/ccd/sn.html}}: $$ \textrm{S/N} =(10^{-\frac{\textrm{emag}}{2.5}}-1)^{-1}       $$

where ``emag" is the magnitude error of the measurement. The PSF extraction and photometry were obtained using the PythonPhot package \citep{Jones2015}. To perform completeness tests, we injected a flat distribution (in magnitude) of 100 synthetic mock point sources per CCD for the 59 best functioning CCDs. 
The shape of the PSF of the injected sources is defined by a double-Gaussian function whose parameters are measured separately on each CCD, thus the shape of the PSF varies across the field of view.
We injected the same amount of flux at the same location for all the synthetic sources in different cases to compare the behaviour of the pipeline in different situations.
Each individual result was obtained running the {\it Mary} pipeline twice on the same field on two sub-sets of images taken on the same night. The injection of fake sources at random locations leads to a probability of blending with other bright or saturating sources that increases with the depth of the images, making the results of the tests conservative. Nevertheless, the tests do not account for possible failure of the processing of CCDs during the analysis, which mainly depends on the goodness of the calibration of the images.

\begin{figure}
\begin{center}
\includegraphics[width=\linewidth]{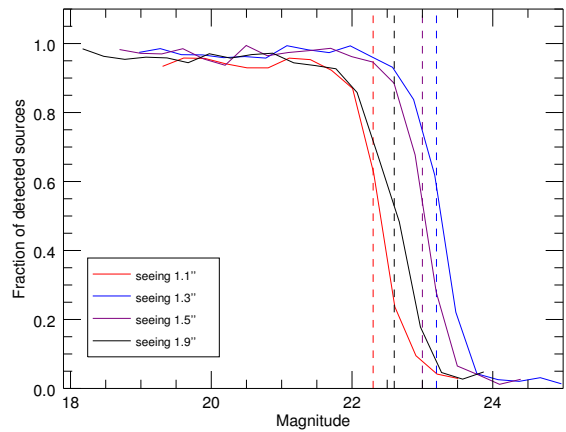}%

\includegraphics[width=\linewidth]{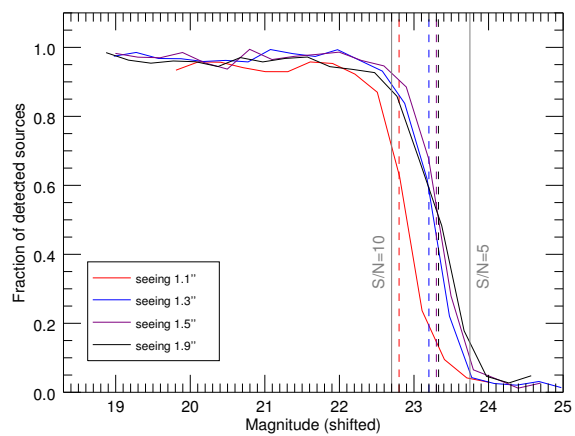}%
\caption{Completeness test run at different observing conditions. The vertical dashed lines intercept the solid-lined curves of the same color where the detection fraction is equal to 0.5, thus where half of the injected sources are recovered by the pipeline. {\it Bottom panel:}  The curves are rigidly shifted to make their 10$\sigma$ limits match the 10$\sigma$ limit of the blue curve (seeing=1.3$''$ ) to facilitate the comparison of the shape of the curves. The S/N=5 vertical line indicates the average magnitude of the 5$\sigma$ limits of the curves, after being shifted.}
\label{plot: completeness combined}
\end{center}
\end{figure}

\begin{figure}
\begin{center}
\includegraphics[width=\linewidth]{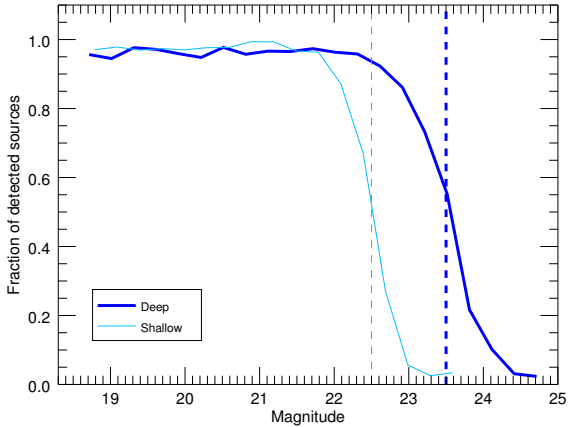}%

\caption{Completeness test performed on stacks of 3 (shallow) and 35 (deep) images taken on 21 December 2015. The vertical dashed lines intercept the solid-lined curve of the same color where the detection fraction is equal to 0.5, thus where half of the injected sources are recovered by the pipeline.}
\label{plot: completeness deep shallow}
\end{center}
\end{figure}

We used ``real" (i.e., not-synthetic) sources to estimate the detection efficiency of the pipeline. Three people independently assessed the transient candidates that {\it Mary} output on the images in Table\,\ref{tab: completeness efficiency} using a template image taken on 17 January 2015, about eleven months before the science images were acquired. The detection efficiency  $\epsilon$ is the ratio of the average number of the ``good" candidates that passed the visual inspection, over the total number of candidates.
We summarise the results of our completeness and efficiency tests in Table\,\ref{tab: completeness efficiency}. In addition, we plot the completeness test results in Fig.\,\ref{plot: completeness combined}-\ref{plot: completeness deep shallow}, where we binned the fraction of detected sources in bins of 0.3\,magnitudes.

We compared the response of {\it Mary} in different seeing conditions by running the pipeline on a set of nine images during each run of the pipeline (Fig.\,\ref{plot: completeness combined}), keeping the same parameters adopted for the timing test (Sec.\,\ref{subsec: timing test}) which set the detection limit at $\sigma_{\textrm{det}} =7.1^{+0.5}_{-0.4}$. We shifted all the curves to make their 10$\sigma$ magnitude limit match the 10$\sigma$ limit of the deepest image (data taken on 20 December 2015, seeing=1.3$''$ ), obtaining the bottom plot of Fig.\,\ref{plot: completeness combined}. The shape of the decaying curve is consistent for three different seeing conditions, which demonstrates the adaptability of the pipeline to different observing situations even without the user acting directly on the control parameters, placing the detection limit at $\sigma_{\textrm{det}} =7.1^{+0.5}_{-0.4} \sigma$ . The outlier red curve (data taken on 22 December 2015, seeing=1.1$''$ ) shows a steeper decay and lower, more scattered values in the part of the curve preceding the decay. In particular, the mean fraction of detected sources with S/N$>$10 is $\mu_{>10\sigma}$=
93.5$\%$, against a mean value of 96.6$\%$ for the other cases. This can be the result of thin clouds present during the observations, and a consequence of the seeing of the science image being better than the seeing of the template image (1.1$''$  vs 1.3$''$ ). This type of problem could be overcome with the implementation of the optimal photometry algorithm \citep{Zackay2016} for the image subtraction.

We tested the response of the {\it Mary} pipeline to a variable number of images to be coadded and searched for transients. Fig.\,\ref{plot: completeness deep shallow} highlights the large gap of 1.0\,magnitude in the detection limit between the two completeness curves obtained by coadding 3 and 35 images, while maintaining the same parameters for the pipeline.

\section{Discussion}
\label{sec:discussion}

The {\it Mary} pipeline was designed to identify optical transient and variable objects in seconds to minutes during the multi-facility observations part of the DWF program.  In this framework, the source identification priority can quickly change, for example in case a LIGO trigger is received and a follow up is performed. Also, observing and data transfer conditions may dramatically change in little time. During DWF observations, a team of astronomers is always ready to modify the parameters of the pipeline according to the priorities of the moment, even if little or no action is required in standard observing conditions. Arbitrary changes may follow decisions to modify the balance between processing speed, completeness down to faint magnitudes, or efficiency in allowing the visual inspection team to confidently identify bright (mag$\sim$21) sources of particular interest. 

For example, a functionality such as the creation of weight maps for the image coaddition can be switched off to save data processing time, especially if images are acquired in a continuous and uniform sequence, improving the performance presented in Sec.\,\ref{subsec: timing test}. If a multi-object spectrograph such as the AAT/AAOmega can provide spectroscopy of hundreds of candidates less than two hours after their detection with DECam, the astronomers can modify the selection criteria to allow a larger completeness. In other cases, multi-object spectrographs and 8m-class telescopes may not be available to follow up DECam candidates due to bad weather, while only smaller telescopes at different observing sites would. In this case, the detection threshold should be moved up in order to facilitate the visual inspection team to focus only on the brightest sources. In general, the users of the pipeline can decide to be more or less conservative in the completeness and candidate selection requirements mainly by modifying the parameters that regulate the source extraction (DETECT\_THRESH\_POS/NEG and DETECT\_MINAREA\_POS/NEG), the shape selection (especially {{\it elmax}), and the bogus rejection parameters ({\it isoareafmin}} and {\it side}). During late-time data analysis on archival images ({\it pipesetup=}NOAO) better completeness and lower efficiency are usually preferred, in order to carefully explore the whole dataset, reducing the risk to miss faint or interesting transients.

In Appendix\,\ref{sec: appendix} we present the default values used for the tests whose results are shown in this paper. These are the values commonly used during DWF observations in standard conditions and are rarely to be modified before or during an observing night. The most sensitive parameters, depending on the FWHM measured for each CCD are automatically computed by the pipeline.

\subsection{Comparison with other pipelines}
\label{subsec:comparison}

The results of the tests of the {\it Mary} pipeline with DECam data to facilitate the comparison with other existing top-level pipelines can be briefly summarised as follows: 
\begin{itemize}
\item Timing:  $\bar{t} \sim 1.1$\,minutes, corresponding to the coaddition of 3 images, optimal for near real-time analysis of the continuous 20\,s exposure images and 20\,s readout time acquired during the DWF observations;  %thus constituting a conservative upper limit for standard observations in which 1-5 images per epoch are acquired;
\item Completeness: we find an average completeness rate for sources with S/N$>$10 and coadding 9 images, where the seeing is comparable to or worse than the template, of  $\bar{\mu}_{>10\sigma} \sim 96.6$ that corresponds to a missed fraction of 3.4$\%$;
\item Detection efficiency: the average value $\bar{\epsilon} = 97.8\%$, obtained with the coaddition of 9 images at different seeing conditions, corresponds to a false positive rate of 2.2$\%$.  
\end{itemize}

The $\bar{t}$ we compute for {\it Mary} is lower than, or comparable with, the timing to run the post-calibration steps of the iPTF pipeline reported by \cite{Cao2016}, specifically 154.3\,s (or 2.57\,minutes) per image. Our $\bar{t}$ is slightly larger than the 60\,s required by the HiTS pipeline to cover the steps between the astrometric calibration and the visual inspection \citep{Forster2016}. The {\it Mary} pipeline %which does $not$ operate any machine learning technique at the moment,
closely approaches the best performing, fully machine-learning-based pipelines in terms of completeness and real/bogus classification ability. In fact, the tests we performed returned a false positive rate of 2.2$\%$ and missed fraction of $3.4\%$ against, for example, the $1\%$ and $5\%$ of iPTF \citep{Cao2016}. These results demonstrate that a pipeline with the design presented in this work is suitable to perform high-quality searches for transients during observing campaigns with DECam such as the DWF program. The {\it Mary} pipeline should produce similar performance for other telescopes and CCDs.

\subsection{Application to other programs and  facilities}
\label{subsec:applications}

The {\it Mary} pipeline was originally designed to work on DECam images and was optimised for the success of DWF observing campaigns. However, the pipeline can be adapted to search for transients in data acquired with different facilities. In this paper we presented {\it Mary} crafted specifically for DWF DECam data, instead of presenting it as a more general pipeline, in part because: 1) small or major necessary changes can not be predicted for each individual telescope; 2) the pipeline is optimised for DECam images, and includes steps (for example the crosstalk rejection) that may not apply to other instruments; 3) the pipeline was designed to run on the Swinburne high performance supercomputer, with large availability of computational resources and parallel processing.

Nevertheless, the {\it Mary} pipeline was used in contexts other than DWF, in order to discover slowly-evolving transients. In particular, {\it Mary} was used during the follow up of Fast Radio Bursts with DECam \citep[Bhandari et al. in prep;][]{Petroff2017}. We also adapted the {\it Mary} pipeline to analyse pre-processed images acquired with the Magellan Fourstar infrared camera \citep{Petroff2017}, discovering tens of variable objects between two consecutive observing nights. Finally, we successfully used {\it Mary} to search for transients in data collected with the 1\,m Zadko telescope, which is among the facilities that make up the DWF follow-up network. This type of analysis was not performed in near real time and dedicated nodes of the Swinburne supercomputer were not required. The {\it Mary} pipeline is currently available upon request, but will be uploaded on GitHub in the future.

\section{Conclusion}
\label{sec:conclusion}

In this paper we presented the {\it Mary} pipeline, designed to discover optical transients in DECam CCD images within seconds to minutes for the DWF program. The pipeline relies on popular astronomical packages to perform key operations and does not involve machine-learning techniques. The performance of such a pipeline (with a false positive rate of $\sim 2.2\%$ and a missed fraction of $\sim 3.4\%$) is similar to the results achieved by the most popular pipelines used by large programs continuously surveying the sky. The {\it Mary} pipeline matches the needs of programs, like DWF, that aim to search for transients over large areas of sky in near real time, with a flexible strategy and variable conditions.

Such a pipeline is crucial to trigger rapid response spectroscopic and imaging observations to study and understand the fast transient Universe. In addition, the {\it Mary} pipeline was successfully used to search for an optical afterglow of a Fast Radio Burst \citep{Petroff2017} with CTIO/DECam and with the FourStar infrared camera on Magellan. 

 We are planning to make the code available to astronomers interested in analysing public DWF data. Finally, pipelines designed on the model of {\it Mary} as presented in this paper will allow more groups of astronomers to search for counterparts to gravitational wave events, Fast Radio Bursts, and new classes of fast transients using small to large facilities.

\begin{acknowledgements}
The authors acknowledge Uros Mestric (SUT) for contributing to the visual inspection of the candidates for this paper, and the group of volunteers who constituted the ``visual inspection team" during the DWF observations in near real time. IA acknowledges Armin Rest (STScI), Phil Cowperthwhite (Harvard), and Frank Masci (IPAC/Caltech) for sharing their knowledge and expertise. 

This research was conducted by the Australian Research Council Centre of Excellence for All-sky Astrophysics (CAASTRO), through project number CE110001020. Research support to IA is provided by the Australian Astronomical Observatory. This work was performed on the gSTAR national facility at Swinburne University of Technology. gSTAR is funded by Swinburne and the Australian Government’s Education Investment Fund. This research was funded, in part, by the Australian Research Council Future Fellowship grant FT130101219 and by the Australian Research Council Centre of Excellence for Gravitational Wave Discovery (OzGrav) through project number CE170100004.
\end{acknowledgements}

\bibliographystyle{apj}
\begin{small}
\bibliography{../../../References/references}
\end{small}

\onecolumn

\begin{center}
\begin{threeparttable}
\caption{Date (UTC, YYMMDD), seeing (arcsec), number of coadded images, and magnitude limit ($g$-band, 5$\sigma$) of the science images used during the completeness ($\mu$) and efficiency ($\epsilon$) tests. We compute $\mu$ as the average number of mock sources with S/N$>10\sigma$ recovered by the pipeline. We base the estimate of $\epsilon$ as the ratio between the number of candidates that pass the visual inspection and the total number of candidates. }
\begin{tabular}{ccccccccc}

\hline\hline
Date & Seeing & N & $\sigma_{\textrm{det}} $& mag$_{\textrm{det}} $ &  mag$_{5\sigma} $ &  mag$_{10\sigma} $ & $\mu_{>10\sigma}$ & $\epsilon$ \\
\hline%
 151219 & 1.7$''$ & 9 & 7.0& 22.6 & 23.0 &22.0& 95.7$\%$ & 98.2$\%$\\
 151220 & 1.3$''$  & 9 & 7.5 & 23.2 & 23.7&22.7&97.0$\%$ & 99.0$\%$\\
 151221\tnote{\textdagger} & 1.5$''$  & 3&7.6 &22.5& 23.1 &22.2&96.7$\%$ & 97.1$\%$\\
 151221 & & 9 &  6.7& 22.4 & 23.4 &22.4 &97.0$\%$ & 96.2$\%$\\
 151221\tnote{\textdagger}  & & 35 &6.3 &23.5& 23.8&22.8&96.0$\%$ & 98.3$\%$\\
 151222 & 1.1$''$  & 9 & 9.5 & 22.3 & 23.0 & 22.2 & 93.5$\%$ & 97.9$\%$\\
\hline\hline
\label{tab: completeness efficiency}
\end{tabular}

 \begin{tablenotes}
 \begin{small}
 \item[\textdagger] Used to compare the performance on shallow and deep images. 
 \end{small}
  \end{tablenotes}
\end{threeparttable}
\end{center}

%\twocolumn

\begin{appendix}
\section{The {\it Mary} parameters}
\label{sec: appendix}

%Table of the parameters.
\begin{longtable}{@{}C{4.0cm} C{1cm} L{8.0cm} @{}}
\caption{The majority of parameters that regulate the {\it Mary} pipeline must be set before starting the analysis of the images from a given observing night (S), or are automatically computed (A). In particular, most of the automatically computed parameters rely on the estimation of the average FWHM of each CCD described in Sec.\,\ref{secsub:science and ref images}. No manual, ``on the fly" intervention is required for the user, while {\it Mary} allows the user to tweak any of the parameters at any time during an observing campaign. In particular, the {\it sequencenumber} parameter, that can uniquely identify the current data processing, was manually (M) modified for each set of images during the past DWF observations in near real time. Default values are indicated for sensitive parameters, as used for the tests reported in this paper, which represent a standard setup during DWF observations.}\\
\hline\hline
Parameter & S/A/M & Description \\
\hline
\multicolumn{3}{l}{\,\,\,\,\,\,    PIPELINE SETUP}\\
\hline
{\it pipesetup} & S & Choose to use science images acquired in near real time (RT) of processed with the NOAO pipeline (NOAO). \\
{\it path\_original} & S & Path where the images processed with the NOAO pipeline are stored.\\
{\it wmapchoice} & S & Choose to create and use weight maps (YES$/$NO, default=YES). \\
{\it checkwcs} & S &  Choose to proceed only in case of successful WCS calibration with PhotPipe (YES/NO, used only with {\it pipesetup}=RT, default=NO).\\
{\it extnum} & S & Number of extensions of the FITS files (default=60).\\
{\it walltime} & S & Maximum ``wall" time for the processing on the Swinburne Green II supercomputer. \\
{\it list\_images\_sci} & S & Path to the list of science images to process. \\
{\it list\_images\_temp} & S &Path to the list of template images to process (only if {\it useoldtemplate}={\it usemarytemplate}=NO). \\
{\it sequencenumber} & M$/$A & Code associated to the {\it Mary} run. \\
{\it field} & S & Name of the observed field. \\ 
{\it date} & S & Date of the observations. \\
{\it filter} & S & Filter used during the observations.\\

\hline
\multicolumn{3}{l}{\,\,\,\,\,    TEMPLATE IMAGES}\\
\hline
{\it datetemp} & S & Date of the template image(s) (only if {\it useoldtemplate}={\it usemarytemplate}=NO ). \\
{\it useoldtemplate} & S & Choose to use a full-field, well-processed image as template (YES$/$NO). \\
{\it oldtemp} & S & Specify the path to the template image (only if {\it useoldtemplate}=YES). \\
{\it usemarytemplate} & S & Coose to use the same template used for another {\it Mary} run (YES$/$NO). \\ 
{\it otherdate} & S & Date of the {\it Mary} run to copy the template from (default={\it datetemp}). \\
{\it othersequncenumber} & S &  Code of the {\it Mary} run to copy the template from. \\
{\it otherfield} & S/A &  Target field of the {\it Mary} run to copy the template from (default={\it field}).\\
{\it otherfilter} & S/A &  Filter used in the {\it Mary} run to copy the template from (default={\it filter}). \\

\hline
\multicolumn{3}{l}{SEXTRACTOR AND HOTPANTS }\\
\hline
{\it DETECT\_MINAREA\_POS} & S & SExtractor DETECT\_MINAREA parameter for the subtracted image (default=8).   \\   
{\it THRESH\_TYPE\_POS} & S & SExtractor THRESH\_TYPE parameter for the subtracted image (default=RELATIVE). \\              
{\it DETECT\_THRESH\_POS }  & S & SExtractor DETECT\_THRESH parameter for the subtracted image (default=1.8).  \\    
{\it ANALYSIS\_THRESH\_POS}  & S &  SExtractor ANALYSIS\_THRESH parameter for the subtracted image (default=1.8). \\
{\it DETECT\_MINAREA\_NEG} & S &SExtractor DETECT\_MINAREA parameter for the inverse of the subtracted image (default=8).\\      
{\it THRESH\_TYPE\_NEG} & S & SExtractor THRESH\_TYPE parameter for the inverse of the subtracted image (default=RELATIVE). \\   
{\it DETECT\_THRESH\_NEG}  & S & SExtractor DETECT\_THRESH parameter for the inverse of the subtracted image (default=1.3). \\    
{\it ANALYSIS\_THRESH\_NEG} & S & SExtractor ANALYSIS\_THRESH parameter for the inverse of the subtracted image (default=1.3). \\
{\it ilsigma} & S & Number of standard deviations from the background value at which the good pixel lower limit is set in HOTPANTS (default=5). \\
{\it tlsigma} & S & Number of standard deviations from the background value at which the good pixel upper limit is set in HOTPANTS (default=5). \\
{\it ker} & A & Convolution kernel half width (default=2.5$\times$\,{\it fwhmsub}). \\

\hline
IMAGE & & \\
\hline

{\it xdim} & S/A & Number of pixels in the X axis of the aligned images (default = 3800). \\
{\it ydim} & S/A & Number of pixels in the Y axis of the aligned images (default = 1800). \\
{\it fixoldchoice} & S & Choose to manually define the size of the aligned images (with {\it xmin, ymin}) even when {\it useoldtemplate}=YES.\\

\hline
\multicolumn{3}{l}{\,\,\,\,\,   SELECTION OF THE CANDIDATES }\\ 
\hline
%{\it choice} & & \\
{\it elmax} & S & Maximum elongation allowed, computed with SExtractor (ELONGATION output key, default=1.8). \\
{\it fwhmtemp} & A & FWHM of the template image.  \\
{\it fwhmsci} & A & FWHM of the science image. \\
{\it fwhmsub} & A & FWHM of the subtracted image.  \\
{\it scale} & S &Pixel scale (arcsec$/$pix, default=0.263).\\
{\it seeing} & A & Seeing of the science image (in arcsec, fwhmsci*scale). \\

{\it isoareafmin} & S & Minimum value allowed for the filtered isophotal area of the sources, computed with SExtractor (ISOAREAF\_IMAGE, default=20). \\
{\it side} & A & Side of the box for the rejection of sources with a counterpart in the inverse of the subtracted image (default=2.0*{fwhmsub}). \\

\hline
\multicolumn{3}{l}{\,\,\,\,\,   MACHINE LEARNING CLASSIFIER}\\
\hline 
{\it mlchoice} & S & Choose whether to use the machine learning classifier to remove CCD artifacts (YES/NO). \\
{\it classthresh} & S & Classification threshold for the machine learning output (between 0 and 1, default=0.5). \\

\hline
\multicolumn{3}{l}{\,\,\,\,\,  CROSSTALK AND CATALOGS}\\
\hline

{\it crosschoice} & S & Choose whether to keep or eliminate those candidates flagged as possible crosstalk (ELIMINATE$/$KEEP). \\
{\it satlevelchoice} & S & Choose whether to use a fixed value (in ADU) above which sources generate crosstalk effects, instead of using the value of the SATURATE header keyword (YES$/$NO). \\
{\it fixsatlevel} & S & Fixed value (in ADU) above which sources are considered to generate crosstalk effects (used only if {\it satlevelchoice}=YES, default=10000).\\

{\it saturchoice} & S & Choose whether to keep or eliminate those candidates flagged as saturating (ELIMINATE$/$KEEP).  \\
{\it maxsources} & S & Maximum number of sources allowed per CCD; an error message is generated if the number of sources is greater than {\it maxsources}\\
{\it querychoice} & S & Choose whether to query the catalogs online or their downloaded version (O$/$D). \\
{\it pathdownloadedGSC} & S/A & Path to the downloaded GSC catalog (only if {\it querychoice}=D).  \\
{\it radiusGSC} & A & Search radius for the GSC catalog, automatically defined as $\sqrt{seeing^2 + u^2}$, where {\it u} is the angular resolution of the GSC catalogue ($u=1''$).  \\
%{\it brightmag} & S & ??? \\
\hline
RANKING & & \\
\hline
{\it nightsarray} & S & Array of dates to be accounted for the ranking of the candidates. \\
{\it matchradius} & S & Radius (in arcsec) to match detections at different epochs. \\

\hline
 PHOTOMETRY & & \\
\hline
 
{\it photochoice} & S & Choose to compute the zero point of the images. \\
{\it catalogcalib} & S & Catalog to use to calibrate the zero point of the images (GAIA$/$USNO-B1)\\
{\it path\_star} & S & Path to the stars pre-selected to perform the zero-point calibration. \\
{\it zpt} & S & Nominal zeropoint to be used if  {\it photochoice}=NO (default=25).\\

\hline
 VISUALISATION & & \\
\hline
{\it radcir\_i} &S & radius (pixels) of the circular regions centred on the candidates (default=12\,pixels).  \\
{\it radcir\_w} & S & radius (arcsec) of the circular regions centred on the candidates (default=8\,arcsec).\\
{\it sidestamp} & S & Measure of the side of the postage stamp centred on the candidate (in pixel, default=120). \\

\hline\hline
\label{tabpar}
\end{longtable}

\end{appendix}

\end{document}